\documentclass[12pt]{article}
\usepackage{amsmath,latexsym,amssymb,epsfig,psfrag}
\usepackage{axodraw}

\def\ltsim{\lower3pt\hbox{$\, \buildrel < \over \sim \, $}}  
\def\gtsim{\lower3pt\hbox{$\, \buildrel > \over \sim \, $}}  
\newcommand{\be}{\begin{equation}}  
\newcommand{\ee}{\end{equation}}  
\def\ga{\mathrel{\raise.3ex\hbox{$>$\kern-.75em\lower1ex\hbox{$\sim$}}}}  
\def\la{\mathrel{\raise.3ex\hbox{$<$\kern-.75em\lower1ex\hbox{$\sim$}}}}

\makeatletter   
\@addtoreset{equation}{section}   
\makeatother

\def\simlt{\stackrel{<}{{}_\sim}}
\def\simgt{\stackrel{>}{{}_\sim}}  

\begin{document}
\baselineskip=16pt   

\begin{titlepage}  

\vskip 2cm
\begin{flushright}
{\bf UAB-FT-581}
\end{flushright}
\begin{center}  
\vspace{0.5cm} \Large {\sc Supersymmetry and Electroweak 
\\ Breaking in the Interval}
\vspace*{5mm}  
\normalsize  
  
{\bf 
D.~Diego~\footnote{diego@ifae.es},
G.v.~Gersdorff~\footnote{gero@pha.jhu.edu} 
and
M.~Quir\'os~\footnote{quiros@ifae.es} 
}   

\smallskip   
\medskip   
\it{$^{1,\,3}$~Theoretical Physics Group, IFAE}\\ 
\it{E-08193 Bellaterra (Barcelona), Spain}

\smallskip   
\medskip   
\it{$^{2}$~Dept. of Physics and Astronomy}\\ 
\it{Johns Hopkins University, Baltimore, MD 21218}

\smallskip     
\medskip
\it{$^3$~Instituci\'o Catalana de Recerca i Estudis Avan\c{c}ats (ICREA)}

\vskip0.6in \end{center}  
   
\centerline{\large\bf Abstract}  

\noindent

Hypermultiplets are considered in the five-dimensional interval where
all fields are continuous and the boundary conditions are dynamically
obtained from the action principle. The orbifold boundary conditions
are obtained as particular cases. We can interpret the Scherk-Schwarz
supersymmetry breaking as a misalignment of boundary conditions while
a new source of supersymmetry breaking corresponding to a mismatch of
different boundary parameters is identified. The latter can be viewed
as coming from boundary supersymmetry breaking masses for hyperscalars
and the nature of the corresponding supersymmetry breaking parameter
is analyzed. For some regions of the parameter space where
supersymmetry is broken (either by Scherk-Schwarz boundary conditions
or by boundary hyperscalar masses) electroweak symmetry breaking can
be triggered at the tree level.

\vspace*{2mm}   
  
\end{titlepage}  
  
\section{\sc Introduction}
The existence of extra dimensions is a general prediction of
fundamental (string) theories that aim to unify all interactions,
including gravity, and provide a consistent quantum description of
them. If the radii of these extra dimensions is as large as the 1/TeV
scale~\cite{Antoniadis:1990ew}, matter can propage in the bulk and the
very existence of extra dimensions can provide new mechanisms for
supersymmetry and electroweak breaking~\cite{Quiros:2003gg}.  In five
and six-dimensional theories gauge bosons are located in vector
multiplets and matter and Higgs bosons in hypermultiplets. While
supersymmetry breaking should be felt primarily by $SU(2)_R$ doublets
(fermions in vector multiplets and bosons in hypermultiplets),
electroweak breaking by the conventional Higgs mechanims concerns the
bosons of the Higgs hypermultiplet. Therefore for a better
understanding of supersymmetry and electroweak breaking we should
consider mainly the system of Higgs hypermultiplets propagating in the
bulk. In this paper we will study propagation of hypermultiplets in
five dimensions (5D).

Propagation of matter in the bulk of the fifth dimension has been
widely considered in the past~\cite{PQ,Barbieri:2000vh}. The 5D
space-time, with coordinates $(x^\mu,y)$, is often constructed as the
orbifold $S^1/\mathbb Z_2$, where points on the circle of radius
$R$~\footnote{We will work from here on, unless explicitly stated, in
  units where $R\equiv 1$.} related by the reflection of the fifth
coordinate $y\to -y$ are identified. In this approach (``upstairs''
approach) fields are classified according to the $\mathbb Z_2$ parity
and their boundary conditions (BC's) at the fixed points imposed. In
the orbifold approach the space is singular at the orbifold fixed
points $y=0,\pi$, it has no boundaries and the fields satisfy the
circle periodicity and the orbifold parity. In this approach mass
terms localized at the fixed points can trigger supersymmetry breaking
and make the fermionic wave functions to make discontinuous jumps at
them~\cite{Bagger:2001ep,Delgado:2002xf}.

An alternative approach is working in the fundamental region of the
orbifold $[0,\pi]$ and giving up the rigid orbifold
BC's~\cite{Csaki:2003dt}. In this approach (``downstairs'' approach)
fields have no defined parity and BC's are dynamically determined by
the action principle~\cite{vonGersdorff:2004eq}. The space is not
singular but has boundaries at $y=0,\pi$ and wave functions are
continuous even in the presence of boundary mass terms. In particular
the propagation of gauge fermions in the interval has been considered
in Ref.~\cite{vonGersdorff:2004eq} where the Scherk-Schwarz (SS)
supersymmetry breaking~\cite{Scherk:1978ta} was interpreted as
misalignment of BC's at the two boundaries that departure from
supersymmetry in boundary parameters.

In this paper we will consider propagation of hypermultiplets in the
interval. We will construct a globally supersymmetric action and
identify possible sources of supersymmetry breaking corresponding to
departure from supersymmetry of boundary parameters. New patterns of
supersymmetry breaking will arise that can trigger electroweak
breaking at the tree-level.  The structure of the paper is as follows.
In section~\ref{hyper} the general supersymmetric formalism for
hypermultiplets in the interval will be worked out. The equations of
motion for hyperscalars and hyperfermions will be solved in
section~\ref{eom} where mass eigenvalues and eigenfunctions are
explicitly obtained.  In particular the conditions for supersymmetry
will be established while supersymmetry breaking by boundary
conditions will be considered in detail in
section~\ref{supersymmetry}. A comparison with the orbifold approach
will be done in section~\ref{orbifold}. The nature of supersymmetry
breaking by boundary hyperscalar masses is clarified in
section~\ref{susybk} where a simple toy model is constructed based on
a $U(1)$ gauge theory under which hypermultiplets transform.  The
problem of embedding $SU(2)_L\otimes U(1)_Y$ in the interval will be
considered in section~\ref{ewsb} where the interface between
supersymmetry and electroweak breaking will be studied, including the
tree-level prediction for the Higgs mass.  Finally
section~\ref{conclusions} contains our conclusions and some
(technical) usuful identities are presented in appendix A.

\section{\sc Hypermultiplets in the interval}
\label{hyper}
In this section we will consider the formalism for a single
hypermultiplet propagating in the interval. There are two equivalent
approaches. One is to consider the fields in the hypermultiplet as
complex and unconstrained fields: it is the so-called complex
hypermultiplet~\cite{Breitenlohner:1981sm}. The second approach is to
introduce an $SU(2)_H$ index on the hypermultiplet
fields~\footnote{The subscript $i$ transforms as a doublet under the
group $SU(2)_R$.}
\be
\mathbb
H^\alpha=(\Phi_i,\Psi,F_i)^\alpha ,
\label{hp}
\ee
that transforms as a doublet, and to introduce on the fields the
reality constraint~\cite{deWit:1984px}
\be
\bar\Phi^i_\alpha\equiv(\Phi_i^\alpha)^*=
\epsilon^{ij}\epsilon_{\alpha\beta}\Phi^{\beta}_j
\label{reality}
\ee
The auxiliary fields obey the same constraint, while the hyperfermions
now obey a symplectic Majorana constraint with respect to the new
$SU(2)_H$~\footnote{The convention is such that
$\epsilon^{12}=\epsilon_{12}=+1$.}
\be
\bar\Psi_\alpha\equiv (\Psi_\alpha)^\dagger\gamma^0=
\epsilon_{\alpha\beta}(\Psi^\beta)^TC\,,
\label{symplectic}
\ee
where $C$ is the 5D charge conjugation matrix.  In the following we
will use real hypermultiplets and conventions and notations are those
of Ref.~\cite{Zucker:2003qv}. In principle we always could explicitely
solve the reality constraints to obtain the standard complex
hypermultiplet, but we will find it useful to express our results in
terms of the real fields. It is important to realize that the doublet
of real fields describes the same degrees of freedom as one complex
hypermultiplet.

We will consider the total action $\mathcal S=\mathcal S_{\rm
bk}+\mathcal S_{\rm bd}$ as the sum of a bulk ($\mathcal S_{\rm
bk}=\mathcal S_{\rm bk}^0+\mathcal S_{\rm bk}^{\rm m}$) and a boundary
($\mathcal S_{\rm bd}$) term, as
\begin{align}
\mathcal S_{\rm bk}^0=&\int_{\cal M}\left( -\frac{1}{2}\bar\Phi\;
\partial^2 \Phi + \frac{i}{2} \bar \Psi\gamma^M\partial_M \Psi+2\bar F
F\right)\label{actionbk0}\\ \mathcal S_{\rm bk}^{\rm m}=&
\int_\mathcal M \left(2i\bar F \mathcal M \Phi + \frac{1}{2}\bar \Psi
\mathcal M \Psi\right)\label{actionbkm}\\ \mathcal S_{\rm
bd}=&\int_{\cal \partial M} \left(\frac{1}{4}\bar \Psi S\Psi
+\frac{1}{4}(\bar \Phi R \Phi)'+\frac{1}{4} \bar\Phi
N(-1+R)\Phi\right)
\label{actionbd}
\end{align}
where $\mathcal S^{\rm m}_{\rm bk}$ in (\ref{actionbkm}) is a
supersymmetric bulk mass action and for simplicity we are using an
indexless notation. $S$ and $\mathcal M$ are matrices in the $SU(2)_H$
indices, $R$ has matrix indices in both $SU(2)_H$ and $SU(2)_R$ and
$N$ is a real number~\footnote{It is understood that $S$, $R$ and $N$
can take on different values at the two branes, i.e.~the usual index
$f=0,\pi$ is suppressed here.}. $S$ and $R$ are dimensionless matrices
and $N$ and $\mathcal M$ have dimension of mass. We will show
in subsection~\ref{susy} that this action is supersymmetric.  In the next
section we will first derive the BC's resulting from the
action~(\ref{actionbk0}), (\ref{actionbkm}) and (\ref{actionbd}).

\subsection{\sc The boundary conditions}
\label{bc}

We take $S$, $\mathcal M$ and $R$ to be hermitian. In order for
the action ~(\ref{actionbk0}), (\ref{actionbkm}) and (\ref{actionbd})
to be real, the reality constraints imply that $S$, $\mathcal M$ and
$R$ satisfy:
\be
S_\alpha^\beta=\epsilon_{\alpha\gamma}\epsilon^{\beta\delta}S_{\delta}^{\gamma}
\label{realityS}
\ee
\be \mathcal
M_\alpha^\beta=\epsilon_{\alpha\gamma}\epsilon^{\beta\delta}\mathcal
M_{\delta}^{\gamma}
\label{realityM}
\ee
\be
R_{i\alpha}^{j \beta}=\epsilon_{ik}\epsilon^{jl}
\epsilon_{\alpha\gamma}\epsilon^{\beta\delta}R_{k\delta}^{l\gamma}
\label{realityR}
\ee
With this choice, all terms in the above action are real without
partial integration because of the reality constraints.  We now make
the ansatz~\footnote{The choice $R=T\otimes S$ is motivated by the
fact that the supersymmetry transformation laws (see
subsection~\ref{susy}) make sense.  The boundary conditions we will find
are the same on both sides of the transformation laws provided
$\epsilon$ fulfills
$
(1+i\gamma^5 T)\epsilon=0.
$
} $R=T\otimes S$ where $T$ acts on $SU(2)_R$ only, i.e.~$T^i_j$. Then
all conditions on $R$, $S$ and $\mathcal M$ can be formulated in
matrix notation as
\be \mathcal M^\dagger=\mathcal M,\qquad \mathcal M^T=-\sigma^2
\mathcal M \sigma^2
\label{condM}
\ee
\be
S^\dagger=S,\qquad S^T=-\sigma^2 S \sigma^2
\label{condS}
\ee
\be
T^{\dagger}=T,\qquad T^{T}=-\sigma^2 T \sigma^2
\label{condT}
\ee
The solution to these constraints are
\be \mathcal M=M \,\vec p\cdot\vec\sigma, \qquad S=\vec
s\cdot\vec\sigma, \qquad T=\vec t\cdot\vec\sigma \label{solcon}\ee
where $\vec s$, $\vec p$ and $\vec t$ are real dimensionless vectors,
$M$ is a mass parameter and $\vec p$ a unit vector. All calculations can now be
performed without writing explicit indices by use of the identities
presented in the Appendix.

Variation of the action gives a bulk and a boundary term $\delta
\mathcal S=\delta_{\rm bk}\mathcal S+\delta_{\rm bd}\mathcal S$ the
latter coming from the partial integration of the variation of the
bulk action and from the variation of the boundary action:
\be \delta_{\rm bd} \mathcal S=\frac{1}{2}\int_{\partial \mathcal M}\left[
\delta \bar \Psi(i\gamma^5+S)\Psi
+\delta\bar\Phi(-1+ R)\Phi' +\delta\bar\Phi'(1+R)\Phi+\delta\bar\Phi\,
N(-1+R)\Phi\right].
\label{variation}
\ee
The BC's resulting from this are
\be
\left(1+i\gamma^5 S\right)\Psi=0,\qquad \bar\Psi\left(1- i\gamma^5 S\right)=0 
\label{BCfermion}
\ee
\be
\left(1+R\right)\Phi=0,\qquad \bar \Phi \left(1+R\right)=0
\label{BCboson}
\ee
\be (-1+R)\left[\Phi'+N\Phi\right]=0,\qquad
\left[\bar\Phi'+N\bar\Phi\right](-1+R)=0
\label{BCboson'}
\ee
We find that for consistent fermionic BC's $\vec s$ has to be a unit
vector~\cite{vonGersdorff:2004eq}.  Furthermore,
Eqs.~(\ref{BCboson})--(\ref{BCboson'}) give rise to eight real BC's at
each brane, whereas we need only four. Thus the bosonic system is
clearly overdetermined unless the $8\times 8$ matrix
\be \left(\begin{array}{cc} 0&1+R\\ -1+R&N(-1+R)
\end{array}\right).
\label{matrix}
\ee
is singular. Its determinant is given by
$
(1-\vec s^{\,2}\vec t^{\;2})^4,
$
which vanishes if $\vec t$ is a unit vector. In fact, it is easy to
see that in this case $(1+R)/2$ and $(1-R)/2$ form mutually orthogonal
projectors on two-dimensional subspaces, and hence we reduce the
number of independent BC's on each brane down to four.

Standard orbifold BC's are obtained as
particular cases by taking $S_f=s_f\, \sigma_3$, $T_f=t_f\,
\sigma_3$, $N=0$. The (independent) parity eigenstates are then given
by $\varphi=\Phi_1^1$, $\varphi^c=\Phi_1^2$, $\psi_L=\Psi_L^1$,
$\psi_R=\Psi_R^1$ and their parities are
\begin{eqnarray*}
\varphi(y_f+y)&=&-\,s_f\,t_f\,\varphi(y_f-y)\,,\\
\varphi^c(y_f+y)&=&+\,s_f\,t_f\,\varphi^c(y_f-y)\,,\\
\psi_L(y_f+y)&=&-\,s_f\,\psi_L(y_f-y)\,,\\
\psi_R(y_f+y)&=&+\,s_f\,\psi_R(y_f-y)\,.
\end{eqnarray*}
However, our formalism allows for more general BC's. As we will see,
it can produce SS-twists in both $SU(2)_R$ as well as $SU(2)_H$ space
with SS parameter given by the angle between the vectors $\vec t_0$,
$\vec t_\pi$ and $\vec s_0$, $\vec s_\pi$ respectively. Furthermore,
we can have mixed BC's for bosons parametrized by the masses $N_f$. A
more detailed and general comparison with the orbifold approach will
be done in section~\ref{orbifold}.

\subsection{\sc Supersymmetry of the action}
\label{susy}
We now want to show that the action (\ref{actionbk0})-(\ref{actionbd})
is indeed supersymmetric.  The transformation laws are given by
\begin{eqnarray}
\delta \Phi_i^\alpha &=&
  i \bar\epsilon_i \Psi^\alpha\nonumber\\
\delta \Psi^\alpha&=&
  -\gamma^M\epsilon^i\partial_M \Phi_i^\alpha+2\epsilon^iF_i^\alpha\nonumber\\
\delta F_i^\alpha&=&
  -\frac{i}{2}\bar\epsilon_i\gamma^M\partial_M\Psi^\alpha
\label{susytrans}
\end{eqnarray}
First consider the bulk part,
Eqs.~(\ref{actionbk0})-(\ref{actionbkm}). Under supersymmetry the
Lagrangian varies into a total derivative which leaves a brane
variation given by
\begin{equation}
\delta_\epsilon \mathcal S_{\rm bk} =\int_{\cal \partial M}\left( -
\bar F\bar\epsilon(i\gamma^5)\Psi
+\frac{1}{2}\bar\Phi\bar\epsilon\gamma^\mu\partial_\mu(i\gamma^5)\Psi
+\frac{i}{2}\bar\Phi\bar\epsilon\Psi'-\bar\epsilon \bar\Phi
\mathcal M \gamma^5\Psi\right)
\label{bulkvariation}
\end{equation}
Now consider the variation of the boundary action, Eq.~(\ref{actionbd}).
\begin{align}
\delta_\epsilon \mathcal S_{\rm bd}=& \int_{\cal \partial M} \left(
-\frac{1}{2}\bar\epsilon\gamma^\mu\partial_\mu\bar\Phi S\Psi
-\frac{1}{2}\bar\epsilon\gamma^5\bar\Phi' S\Psi\right.\nonumber\\ +&
\bar\epsilon\bar F S \Psi +\frac{i}{2}\bar\Phi' R\bar\epsilon \Psi
\left.+\frac{i}{2}\bar\Phi R\bar\epsilon \Psi'
+\frac{i}{2}N\bar\Phi(-1+R)\bar\epsilon\Psi\right)
\label{branevariation}
\end{align}
Discarding a total 4D derivative we can rewrite the sum of
(\ref{bulkvariation}) and (\ref{branevariation}) as
\begin{multline}
\delta_\epsilon \mathcal S=
\int_{\cal \partial M} \left(
\frac{i}{2}\bar\epsilon\bar\Phi(1+R) \Psi' +  
\frac{i}{2}(\bar\Phi'+N\bar\Phi) (-1+R)\bar\epsilon \Psi \right.\\ 
\left.
-\frac{1}{2}\bar\epsilon \gamma^5\bar\Phi\mathcal M(1+i\gamma^5S)\Psi
-i\bar\epsilon\gamma^5\left(\bar F-\frac{i}{2}\bar\Phi
\mathcal M \right)(1-i\gamma^5S) \Psi  \right)
\label{finalvar}
\end{multline}
Using the BC's~(\ref{BCfermion}), (\ref{BCboson}) and (\ref{BCboson'})
the first three terms vanish.  Finally we use the EOM for $F$
\be
F=-\frac{i}{2}\mathcal M\Phi,\qquad \bar F=\frac{i}{2}\bar \Phi \mathcal M
\ee
to deduce that the whole variation is zero.

\section{\sc Equations of motion and the spectrum}
\label{eom}
In this section we will solve the equations of motion (EOM) in the
bulk for the bosonic and fermionic sectors of the hypermultiplet and
impose on the solutions the corresponding BC's on both boundaries. We
will obtain as a result the mass eigenfunctions and eigenvalues for
the different modes.

\subsection{\sc Hyperscalars}
\label{bosons}

We first make a general mode decomposition of bosonic fields as
\be
\Phi_i^\alpha(x,y)=\sum_n f_{i,n}^\alpha(y)\; \phi_n(x)
\label{decompbos}
\ee
where $\phi_n(x)$ is the real 4D mass eigenstate corresponding to the mass
$m_n$~\footnote{From here on and for notational simplicity we will
drop, unless explicitly stated, the subscript $n$ as well as we will
use the compact notation where the $i$ and $\alpha$ indices are
omitted.}. The solution to the EOM arising from the bulk action
(\ref{actionbk0}) and (\ref{actionbkm}) is given by
\be
f(y)=\cos(\Omega y)\, a+ \sin(\Omega y)\, b
\label{solbos}
\ee
where $a_i^\alpha$ and $b_i^\alpha$ are constant matrices and
$\Omega_n=\sqrt{m_n^2-M^2}$.

The BC's at $y=0$, Eqs.~(\ref{BCboson}) and (\ref{BCboson'}) yield
\begin{align}
a=&(1-R_0)\, \varphi\nonumber\\
b=&\left[1+R_0-\frac{N_0}{\Omega}(1-R_0)\right] \varphi
\label{BCbos0}
\end{align}
where $\varphi$ is a constant unconstrained matrix. Imposing the BC's
at $y=\pi$ determines $\varphi$ and gives the discrete eigenvalue
spectrum as a function of $\vec s_f$ and $\vec t_f$.
Defining the angles $\omega$ and $\tilde\omega$ by
\begin{align}
&\vec s_0\cdot\vec s_\pi=\cos(2\pi\tilde\omega)\nonumber\,,\\
&\vec t_0\cdot\vec t_\pi=\cos(2\pi\omega)\,,
\label{anglebos}
\end{align}
the bosonic mass eigenvalues are given as the solutions of the equation
\begin{gather}
A(\omega+\tilde\omega,N_0,N_\pi)A(\omega-\tilde\omega,N_0,N_\pi)=0\nonumber\\
A(\varphi,N_0,N_\pi)=
\sin^2(\pi \varphi)\,
-\,\frac{N_0-N_\pi}{\Omega}\,\tan(\Omega\pi )
-\left[\cos^2(\pi\varphi)   + \, \frac{N_0\,
N_\pi}{\Omega^2}\right]\tan^2(\Omega\pi )
\label{condbos}
\end{gather}
Notice that the parameter $\omega$ describes a Scherk-Schwarz twist
and thus it corresponds to supersymmetry breaking. The parameter
$\tilde\omega$ is a twist in the global $SU(2)_H$ symmetry and amounts
to a supersymmetric mass, as we will see in the next section. As for
the mass parameters $N_{f}$ they can conserve or break supersymmetry
depending on their relation with $M\,\vec p\cdot \vec s_f$ as we will
see. For special values of $(\omega,\tilde\omega)$ the mass formula
becomes a perfect square, indicating a degeneracy in the spectrum.  The
values where this happens are given by $(\omega,0)$,
$(\omega,\frac{1}{2})$, $(0,\tilde\omega)$ and $(\frac{1}{2},\tilde
\omega)$.

We want to close this section by noticing that the condition for the
existence of an exactly massless mode is given by the equation
\be
(n_0-\tau^{-1})(n_\pi+\tau^{-1})=\cos^2\pi(\omega\pm\tilde\omega)(1-\tau^{-2})
\label{hyp}
\ee
where 
\be
n_f=N_f/M
\label{enes}
\ee
and $\tau\equiv \tanh  (M \pi )$. This defines a
hyperbola which divides the ($n_0$,$n_\pi$) plane in regions where the
lightest mode is tachyonic or physical respectively. This will be
discussed in more detail in section~\ref{supersymmetry}.

\subsection{\sc Hyperfermions}
\label{fermions}

We will define the Dirac fermion in the 5D action as
\be
\Psi^\alpha=\left(
\begin{array}{c}
\chi^\alpha\\
\bar\psi^\alpha
\end{array}
\right)
\ee
where $\chi^\alpha$ and $\bar\psi^\alpha$ are Weyl fermions subject to
the reality condition
$\psi_\alpha=\epsilon_{\alpha\beta}\chi^\beta$. We will make the mode
decomposition
\begin{align}
\chi^\alpha(x,y)=& \sum_n f^\alpha_n(y)\chi_n(x)\nonumber\\
\bar\psi^\alpha(x,y)=& \sum_n g^\alpha_n(y)\bar\chi_n(x)
\label{decompfer}
\end{align}
where $(\chi_n(x),\bar\chi_n(x))^T$ is the 4D Majorana spinor with mass
eigenvalue $m_n$.
We now define the vector 
\be
h(y)=\left(
\begin{array}{c}
f(y)\\
g(y)
\end{array}
\right)
\label{espacio}
\ee
where we have dropped the indices $\alpha$, $n$. The bulk EOM
corresponding to the bulk action (\ref{actionbk0}) and
(\ref{actionbkm}) has the solution
\be h(y)=\mathcal U(y)\ h(0);\quad \mathcal U(y)=\cos(\Omega
y)+(im\sigma^2+\mathcal M\sigma^3)\,\frac{\sin(\Omega y)}{\Omega} 
\label{solfer}
\ee
where $\sigma^{2,3}$ are acting on the space of Eq.~(\ref{espacio})
and $\mathcal M$ is acting on $SU(2)_H$ indices.

We now apply the BC's (\ref{BCfermion}) at the two boundaries. In
particular the BC's at $y=0$ imply that
\be
h(0)=\left(1-\sigma^3 S_0\right)\;\tilde h
\ee
and at $y=\pi$
\be V(\pi)\;\tilde h\equiv\left(1+\sigma^3 S_\pi\right)\mathcal
U(\pi)\left(1-\sigma^3 S_0\right)\,\tilde h=0 \ee

The $4\times 4$ matrix $V(\pi)$ has rank $r\leq 2$ because it is
proportional to the projector $(1+\sigma^3 S_\pi)/2$. The existence of
a non-trivial solution requires $r=1$ which provides the constraint
satisfied by the mass eigenvalues.  The result can be expressed in
terms of the angles $\tilde\omega$ defined in Eq.~(\ref{anglebos}) and
$\alpha_f$ defined by 
\be
\vec p\cdot\vec s_f=\cos(2\pi\alpha_f)\equiv
c_f
\label{ces}
\ee
The fermion mass eigenvalues satisfy then the equation
\be 1-\tilde c-2\,(c_0-c_\pi)\frac{M}{\Omega}\tan(\Omega\pi )
-\left[1+\tilde c+2\, c_0\, c_\pi
\frac{M^2}{\Omega^2}\right]\tan^2(\Omega\pi )=0
\label{condfer2}
\ee
where the quantity $\tilde c=\cos(2\pi\tilde\omega)$.
Note that the quantities $c_0$ and $c_\pi$ are not completely independent
but are bound to lie inside an elliptical disk
\be
\frac{(c_0+c_\pi)^2}{\cos^2 \pi\tilde\omega}+
\frac{(c_0-c_\pi)^2}{\sin^2 \pi\tilde\omega}\leq 4\ .
\label{ellipse}
\ee
This condition stems from the fact that the three angles between the
three vectors $\vec s_f$, $\vec p$ are not independent but rather
constrained by triangle inequalities. For instance, if $\tilde c=1$ it
is clear that $0\leq c_0=c_\pi\leq 1$ (in this case the ellipse
actually shrinks to a line).  It will also be convenient to express
the condition Eq.~(\ref{condfer2}) in terms of the function $A$
defined in Eq.~(\ref{condbos}):
\be 
A(\tilde\omega ,c_0 M, c_\pi M)=0\,.
\label{condfer3}
\ee

The bosonic, Eq.~(\ref{condbos}), and fermionic, Eq.~(\ref{condfer3}),
spectra can easily encompass the cases already studied in the
literature. For instance for the particularly simple case where
$N_f=M=0$ the bosonic spectrum provided by Eq.~(\ref{condbos}) is
given by $m_n=n\pm\omega\pm\tilde\omega$ while the fermionic one,
provided by Eq.~(\ref{condfer2}), is given by $m_n=n\pm\tilde\omega$
in agreement with the results of the model studied in Ref.~\cite{PQ}.

For the case $\tilde\omega=0$, $c_0=c_\pi=1$ (i.e.~the three vectors
$\vec s_f$, $\vec p$ aligned) and $N_f=M$ the bosonic spectrum from
Eq.~(\ref{condbos}) is given by the solution of
\be
\sin^2(\pi\omega)=\frac{\Omega^2+M^2}{\Omega^2}\sin^2(\Omega\pi )
\ee
while the fermionic spectrum is given by
$m_n^2=n^2+M^2(1-\delta_{n0})$ in agreement with the results in
Ref.~\cite{Rayner}.  While other cases can be easily studied using the
general equations we will next concentrate in particularly interesting
cases for physics purposes.  In particular, we will examine how
supersymmetry can be broken and how vectorlike fermions can arise.

\section{\sc Supersymmetry breaking}

\label{supersymmetry}

The bosonic spectrum described as the solution of Eq.~(\ref{condbos})
depends on four dimensionless parameters: $\omega$, $\tilde\omega$ and
$n_f$. Similarly the fermionic spectrum described as the
solution of Eq.~(\ref{condfer2}) depends on three parameters:
$\tilde\omega$ and $\alpha_f$. The
parameter $\omega$ is a genuine Scherk-Schwarz supersymmetry breaking
parameter while a particular relation between $n_f$ and $\alpha_f$ can
conserve/break supersymmetry as we will see in this section. Finally
$\tilde\omega$ affects to both bosons and fermions and can play the
role of a supersymmetric mass if the only source of supersymmetry
breaking is the parameter $\omega$ as we have seen in the simple
example described at the end of the previous section.

Comparison between (\ref{condbos}) and (\ref{condfer3}) dictates the
supersymmetric relation between $n_f$ and $c_f=\cos(2
\pi\alpha_f)$. Indeed this is given by
\begin{equation}
n_f=c_f
\label{ecuacion}
\end{equation}
If $n_f$ does not satisfy the relations (\ref{ecuacion}) supersymmetry
is broken. In fact even if the Lagrangian is (on-shell) supersymmetric
the spectrum is not.  This source of supersymmetry breaking can also
be understood as follows.  The BC's
Eqs.~(\ref{BCfermion})--(\ref{BCboson'}) are generally not stable
under the supersymmetry transformations Eq.~(\ref{susytrans}); in
other words the variation of the fields does not fulfill the BC's. As
can easily be shown, the BC's are stable if and only if $n_f=c_f$.  In
summary supersymmetry breaking arises from two different sources: one
is the non-alignment of the vectors $\vec t_f$ (or equivalently
$\omega\neq 0$); another one is the departure from zero of $n_f-c_f$.
In both cases the supersymmetric limit is continuously connected which
suggests that in the locally supersymmetric extension of the action
local supersymmetry might be spontaneously broken. This point is
extremely important and deserves a detailed investigation.

\subsection{\sc Boundary supersymmetry-breaking hyperscalar masses}
\label{susyspectrum}
In order to discriminate between the Scherk-Schwarz mechanism and
other sources of supersymmetry breaking, we will first fix $\omega=0$.

\begin{figure}[hbt]
\centering
\scalebox{0.9}{
\mbox{
\epsfig{file=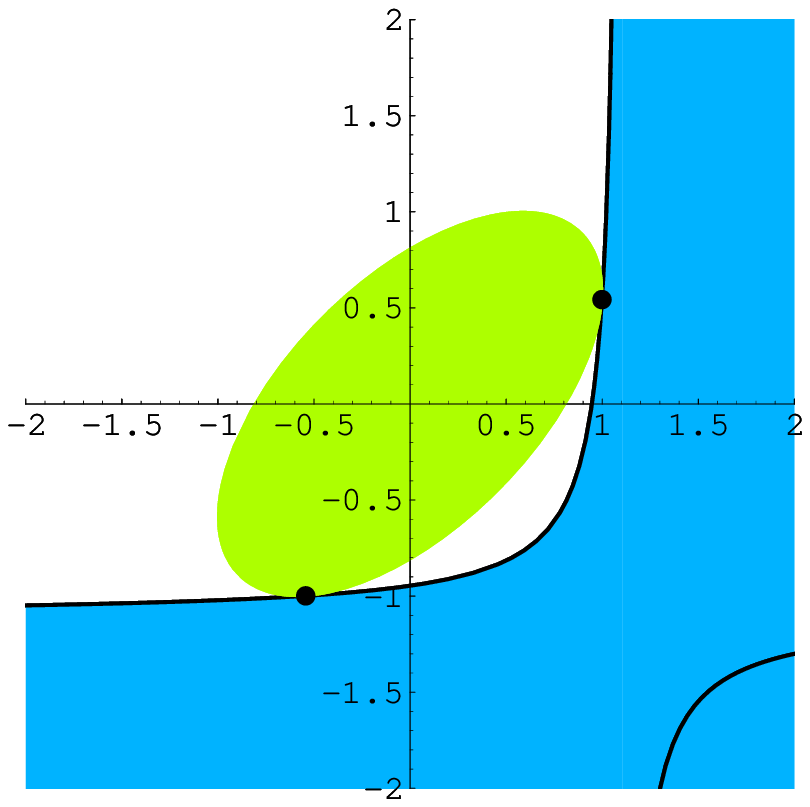}
\hspace{1cm}
\epsfig{file=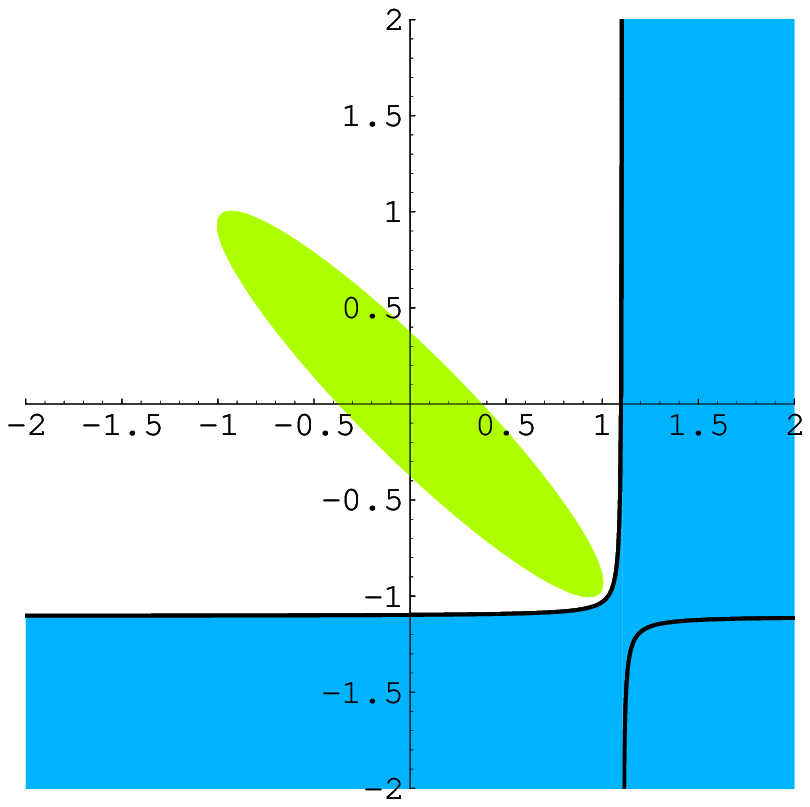}
}}
\caption{\it The hyperbola of massless modes for $\pi MR=1.5$,
$\omega=0$ and $\tilde\omega=0.15$ (left panel) and
$\tilde\omega=0.44$ (right panel) in the plane $(n_0,n_\pi)$.  The
clear region to the upper left has no tachyonic modes, the darkly
shaded (blue) region between to the two branches has one tachyonic
eigenvalue, and the region to the lower right has two tachyonic modes.
The ellipse corresponds to the allowed points in the plane
$(c_0,c_\pi)$. The two dots mark the points where the fermions are
massless.}
\label{regions2}
\end{figure}

As mentioned earlier, the massless bosonic modes lie on a hyperbola in
the $(n_0,n_\pi)$ plane, Eq.~(\ref{hyp}).  It is clear that the mass
squared of one eigenmode changes sign when one crosses this curve. As
can be explicitely checked from Eq.~(\ref{condbos}), at $n_0=n_\pi=0$
there are no tachyonic modes, while at $n_0=n_\pi\gg 1$ there is one
(complex) tachyon with $m^2=-N_0^2$ and at $n_0=-n_\pi\gg 1$ there are
two degenerate tachyons with $m^2=-N_0^2$. It is then easy to identify
three distinct regions with different number of tachyonic
eigenvalues~\footnote{ Some tachyonic spectra will be investigated in
  subsection ~\ref{susybreak} and used for electroweak symmetry
  breaking in section~\ref{ewsb}.}.  For the case $\omega=0$, we
illustrate this situation in Fig.~\ref{regions2} by showing a plot in
the $(n_0,n_\pi)$ plane.
In the same plot we also provide the allowed values
of $c_{0,\pi}$, defined by Eq.~(\ref{ellipse}). The interior of the
ellipse corresponds to the allowed values in the plane $(c_0,c_\pi)$.
The supersymmetric points $n_f=c_f$ are thus also limited to the
inside of the ellipse.  Note that for $\omega=0$ the ellipse cannot
overlap with the shaded region, for this would mean the fermions to
acquire tachyonic masses. However, for certain values of $M$ and
$\tilde \omega$, there are two points where the ellipse is tangent to
the hyperbola, the intersection points corresponding to exactly
massless supersymmetric spectra.  These points are given by
\be
\left(
\begin{array}{cc}c_0\\c_\pi
\end{array}
\right)
=
\tau^{-1} 
\left(
\begin{array}{cc}
\sin^2\pi\tilde\omega\pm \cos\pi\tilde\omega\sqrt{\tau^2 
-\sin^2\pi\tilde\omega}\\
-\sin^2\pi\tilde\omega\pm \cos\pi\tilde\omega\sqrt{\tau^2 
-\sin^2\pi\tilde\omega}
\end{array}
\right)
\ee
and are obviously constrained to 
\be
\tau^2 \geq \sin^2 \pi\tilde\omega\,.
\ee
For other values of $\tau$ and $\tilde\omega$ there is no intersection
of the ellipse with the hyperbola and hence there are no massless
fermions. This is actually the case in the right panel of
Fig.~\ref{regions2}. Notice that there all points $n_f=c_f$ now
correspond to supersymmetric but massive spectra.

The case $\tilde \omega=\frac{1}{2}$ is special.  In fact if the
hypermultiplet $\mathbb H$ transforms non-trivially under the gauge
group the 4D theory might be anomalous if the fermion modes are not
paired to get a Dirac mass. This happens for instance if the
hypermultiplet scalar zero mode is identified with the Higgs field
doublet $H$ in the Standard Model~\footnote{This possibility will be
analyzed in detail in section~\ref{ewsb}.}. A quick glance at
Eq.~(\ref{condfer2}) shows that a sufficient condition for this to
happen is $\tilde\omega=1/2$, i.e.~$\tilde c=-1$.  In that case the
ellipse degenerates to the line $c_0=-c_\pi$ and the spectrum becomes
vectorlike, as Eq.~(\ref{condfer2}) for fermions (Higgsinos) becomes a
perfect square
\be
\left\{
1-c_0\frac{M}{\Omega}\tan(\Omega\pi )
\right\}^2=0
\label{fermvecspec}
\ee
For instance, when $c_0=0$, then the fermionic spectrum is given by
\be \Omega_n=n+\frac{1}{2}\quad \Longrightarrow\quad
m_n^2=M^2+\left(n+\frac{1}{2}\right)^2\,.
\label{fermspec}
\ee
For $c_0=-c_\pi=-1$ the spectrum can be calculated in the large $M$
limit and is given by 
\be
\Omega_n=n+{\cal O}(M^{-1})\quad \Longrightarrow\quad m_n^2=M^2+n^2
+{\cal O}(M^{-1}),\qquad n\geq 1
\label{ferm-1}
\ee
Finally, for the case $c_0=-c_\pi=1$ there is an
exactly massless Dirac fermion in the limit $\tau=1$ ($M\to \infty$).
In fact, it can be shown that for $MR\simgt 1$ there is a light Dirac
fermion with mass
\be
 m = 2 M \exp(-\pi M ).
\label{lightferm}
\ee
Interestingly enough the wavefunctions of the two chirality degrees of
freedom localize towards different branes.

\subsection{\sc Scherk-Schwarz supersymmetry breaking}
\label{susybreak}

\begin{figure}[hbt]
\centering
\scalebox{0.9}{
\mbox{
\epsfig{file=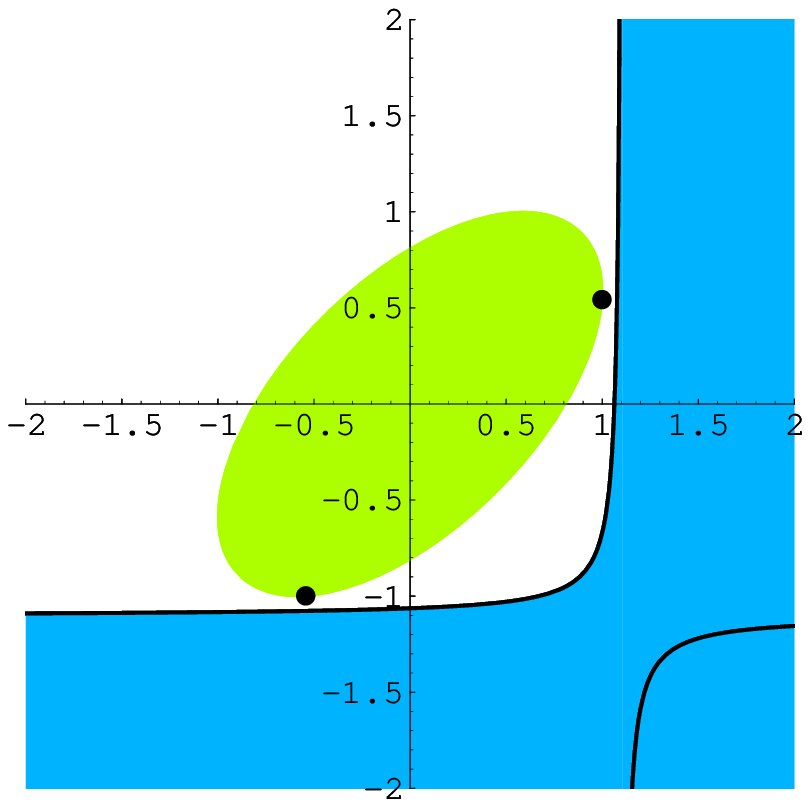}
\hspace{1cm}
\epsfig{file=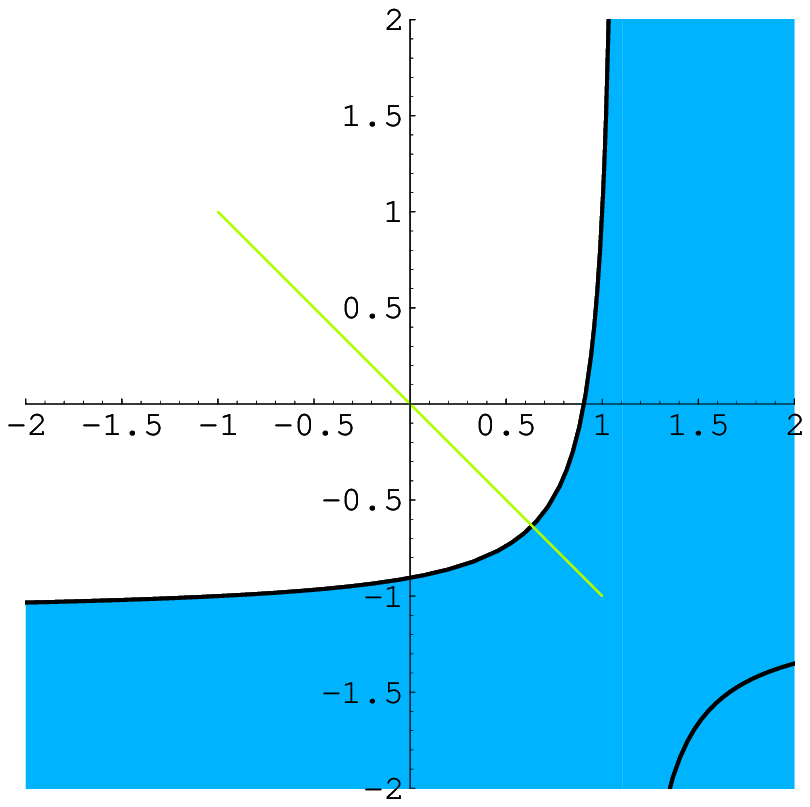}
}}
\caption{\it Same as Fig.~\ref{regions2}, but with $\omega=\frac{1}{2}$ and
$\tilde\omega=0.15$ (left panel) and $\tilde\omega=\frac{1}{2}$ (right
panel.}
\label{regions}
\end{figure}

For $\omega\neq 0 $ the situation changes. Even for $c_f=n_f$ the
theory does provide different spectra for fermions and bosons as
supersymmetry is now broken by the Scherk-Schwarz mechanism.  In
Fig.~\ref{regions} we show the situation in the case
$\omega=\frac{1}{2}$.
In the left panel ($\tilde \omega=0.15$) one can see that for
$n_f=c_f$ bosons always have strictly positive mass-squared, even at
the points where the fermions are massless~\footnote{Notice that the
masslessness condition for fermions is still determined by the
intersection of the ellipse with the $\omega=0$ hyperbola, not shown
in the plot.}. In order to have massless or tachyonic scalars, one has
to move away from points $n_f=c_f$, thereby introducing the new source
of supersymmetry breaking discussed above.  

\begin{figure}[hbt]
\centering
\scalebox{0.9}{
\mbox{
\epsfig{file=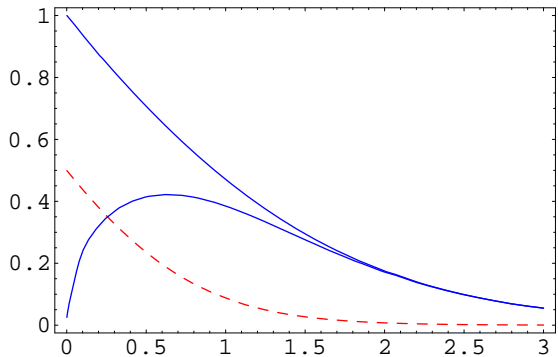}
\hspace{1cm}
}}
\caption{{\it
Light particle spectrum for $\omega=\tilde\omega=\frac{1}{2}$,
$n_0=c_0=1$, $n_\pi=c_\pi=-1$. The plot shows $|m R|$ as a function of
$M R$. The blue (solid) line corresponds to the bosons, the lower
curve being the tachyon.  The red (dashed) is the mass of the lightest
fermion.}
}
\label{light}
\end{figure}

In the right panel of Fig.~\ref{regions} we choose
$\tilde\omega=\frac{1}{2}$. As opposed to the previous case, it is now
possible to have a tachyon at $n_f=c_f$. A particularly interesting
choice is $n_0=c_0=-n_\pi=-c_\pi=1$ where (for $MR\simgt 1$) there are
two light scalars with masses squared 
\be
m^2=\pm 4 M^2 \exp( -\pi MR)
\label{lightbos}
\ee
while the vectorlike fermion zero mode is still given by
Eq.~(\ref{lightferm}). These particular cases were already considered
in Refs.~\cite{Marti,Barbieri:2002uk} in the context of the orbifold
approach~\footnote{The relation to the orbifold approach will be
  clarified in section~\ref{orbifold}.}.  In Fig.~\ref{light} we show
the numerical solution for these masses for general values of $M$.
Notice that for $MR\gg 1$ both bosonic and fermionic masses are
exponentially suppressed with the corresponding phenomenological
troubles. However for values $MR\ll 1$ the exponential suppression
disappears, although the fermion remains lighter than the Higgs boson
for $MR\simgt 0.2$, and correspondingly the present experimental
bounds on charginos can put a lower bound on the Higgs mass in this
class of models. The fact that the zero mode boson is tachyonic and
that supersymmetry is broken by Scherk-Schwarz boundary conditions
provides a priori a very promising class of models of electroweak
symmetry breaking~\footnote{Of course radiative corrections have to be
  taken into account. For $M=0$ this has been done in Ref.~\cite{PQ}
  and it was found in Ref.~\cite{Barbieri:2002uk} that EW symmetry
  breaking can occur as long as the localizing mass term for the top
  hypermultiplet is not too big. Of course the situation changes when
  we slightly localize the Higgs field and allow for a non-vanishing
  value of the parameter $M$. In that case the tachyonic tree level
  mass becomes more and more important when M increases, reaching a
  maximum at around $MR\sim 0.5$, and in this region EW symmetry
  breaking can occur even with a fully localized top.}. In fact
supersymmetry breaking is supersoft (finite) as it is due to global
effects typical of the Scherk-Schwarz breaking, while electroweak
symmetry is accomplished at the tree level and the electroweak
breaking scale (Higgs mass) can be decoupled from the inverse radius
of compactification, which can help to solve the little hierarchy
problem. The wave function of the Higgs is exponentially localized to
one of the interval boundaries, which can help in (partially) solving
the problem of fermion masses while the radion can be stabilized by
the Casimir energy at one or two-loop order as recently
proposed~\cite{radion}. A detailed analysis of these, and other
questions (outside the scope of the present paper) will be considered
elsewhere.

Another particularly interesting supersymmetry breaking case is
$\omega=\frac{1}{2}$ and $N_0=N_\pi\equiv N$ where supersymmetry is
broken by both the SS parameter and by the boundary terms $N_f$. The
bosonic spectrum is now determined by the equation
\be
(\Omega^2+N^2)\;\frac{\tan^2(\Omega\pi)}{\Omega^2}=0
\ee
with solutions
\begin{align}
m_n^2=&\;n^2+M^2 \ (n\neq 0)\nonumber\\
m_0^2=&\,M^2-N^2
\label{bosspec}
\end{align}
The spectrum for the case $M=N=0$ coincides with the $\mathbb
Z_2\times\mathbb Z'_2$ model of Ref.~\cite{Barbieri:2000vh} while the
case with $M=N\neq 0$ is a generalization but still the bosonic zero
mode is massless. The case with $M\neq N$ is a different
generalization where the bosonic zero mode is massive. An interesting
possibility is when $M<N$ in which case the zero mode is a tachyonic
state and can play the role of the Higgs doublet of the Standard Model
and trigger electroweak (and supersymmetry) breaking at the tree
level. In particular the case $M=0$ reproduces the fermionic sector of
the model of Ref.~\cite{Barbieri:2000vh} while for $N\neq 0$ the
bosonic sector has a tachyonic zero mode.

The wave function for the bosonic modes is given, from (\ref{BCbos0}),
as
\be
f_n(y)=\left( \cos(\Omega y)(1-R_0)+\frac{\sin(\Omega
y)}{\Omega}\left[1+R_0-N_0(1-R_0)\right]\right)\varphi
\ee
where $\varphi$ has to be determined from the BC's at $y=\pi $. For
the bosonic sector of the previous case where $R_0=R_\pi$,
$N_0=N_\pi=N$, the wave function of the zero mode, satisfying
$\Omega^2+N^2=0$, is defined by
\be
f_0(y)=e^{-Ny}\varphi\quad {\rm with} \quad (1+R)\varphi=0
\label{wave}
\ee
Two components of $\varphi$ are projected away by the last condition
in (\ref{wave}) while the reality condition implies that only one
independent component is kept. The latter one is fixed by the
normalization condition. Eq.~(\ref{wave}) shows that for $N\neq 0$ the
bosonic zero mode is localized at the $y=0$ ($y=\pi$) boundary for
$N>0$ ($N<0$).

In this case we do not expect supersymmetry breaking to be supersoft
because, on top of the non-vanishing Scherk-Schwarz parameter we have
a departure from the supersymmetric relation (\ref{ecuacion}). On the
other hand, from the phenomenological point of view the Higgsino
masses are much larger than the Higgs mass and present experimental
bounds on chargino masses do not constrain at all the present model.
In fact the nature of the supersymmetry breaking will be clarified in
section \ref{susybk} while some comments about electroweak breaking
for this class of models will be presented in section~\ref{ewsb}.

\section{\sc Comparison with the orbifold approach}
\label{orbifold}
In order to compare the previous formalism with the more usual
orbifold approach, and to also shed light on the nature of the
previously considered supersymmetry breaking, we show in this section
that the same physical theory can be obtained if one considers the
orbifold $S^1/\mathbb Z_2$. We assign the following parities to the
fields
\be \Psi(-y)=i\gamma^5 \sigma_3\Psi(y)\,,\qquad \bar
\Psi(-y)=-\bar\Psi(y)i\gamma^5 \sigma_3\,,
\label{BCfermionorb}
\ee
\be \Phi(-y)=\sigma_3\otimes\sigma_3\Phi(y)\,,\qquad \bar\Phi(-y)=\bar
\Phi(y)\sigma_3\otimes\sigma_3\,,
\label{BCbosonorb}
\ee
\be F(-y)=-\sigma_3\otimes\sigma_3F(y)\,,\qquad \bar F(-y)=-\bar
F(y)\sigma_3\otimes\sigma_3\,.
\label{BCauxorb}
\ee
We also could introduce Scherk-Schwarz twists for the $SU(2)_R$ and
$SU(2)_H$ symmetries. However since the presence of an
$\tilde\omega\neq 0$ parameter amounts to a supersymmetric mass, while
the nature and interpretation of a Scherk-Schwarz twist $\omega\neq 0$
has been widely clarified in the
literature~\cite{Bagger:2001ep,Delgado:2002xf,vonGersdorff:2004eq}, we
will simplify our disscussion in this section by assuming
$\omega=\tilde\omega=0$.  Furthermore, we replace the action given in
Eqs.~(\ref{actionbk0})--(\ref{actionbd}) by
\begin{align}
\mathcal S_{\rm bk}^0=&\int\left( -\frac{1}{2}\bar\Phi\;
\partial^2 \Phi + \frac{i}{2} \bar \Psi\gamma^M \partial_M \Psi+2\bar F
F\right)\,,\label{actionbk0orb}\\ 
\mathcal S_{\rm bk}^{\rm m}=&
\int \left(2i\bar F \mathcal M \Phi + \frac{1}{2}\bar \Psi
\mathcal M \Psi\right)\,,\label{actionbkmorb}\\ 
\mathcal S_{\rm bd}=&\int  
 \biggl(N_0 \delta(y)-N_\pi \delta(y-\pi)\biggr)
 \bar\Phi
\Phi\,.
\label{actionbdorb}
\end{align}
In order to have well-defined parity for the mass terms, we take the
vector $\vec p$ defined in Eq~(\ref{solcon}) to be
\be
p=(p_1,p_2,\epsilon(y)p_3)\ ,
\ee
where $\epsilon(y)$ is the sign-function. Choosing $p_1=p_2=0$ one
reproduces the odd mass terms for hypermultiplets previously
considered in the literature~\cite{Barbieri,Nibb,Marti,Rayner}.  The
boundary mass terms involving the $N_f$ parameters are similar to the
ones encountered in Eq.~(\ref{actionbd}). In fact the boundary
conditions~(\ref{BCbosonorb}) require $R=-\sigma_3\otimes\sigma_3$, so
that by using this in Eq.~(\ref{actionbd}) we find
Eq.~(\ref{actionbdorb}).  The additional factor of 2 comes from the
fact that the support of the delta function on the circle is twice the
one on the interval, while the relative sign of the two boundaries
reflects our convention of taking the orientation of the boundary at
$y=0$ to be negative.  Boundary mass terms --which in the interval
give rise to boundary conditions-- on the orbifold generate jumps for
the profiles of wave functions across the brane. It is easy to
calculate these jumps for the special kind of mass terms of
Eq.~(\ref{actionbdorb}). All fields are continuous except the
$\partial_5$ derivatives of even bosonic fields, which satisfy
\be
(1+\sigma_3\otimes\sigma_3)[\Phi'(0^+)+N_0\Phi(0)]=0\,,
\ee
\be
(1+\sigma_3\otimes\sigma_3)[\Phi'(\pi^-)+N_\pi\Phi(\pi)]=0\,.
\ee
Here we write the matrix $(1+\sigma_3\otimes\sigma_3)$ to project on
the even fields only. The spectrum can now be directly inferred from
subsections~\ref{bosons} and \ref{fermions}. The bosonic one is given
by Eq.~(\ref{condbos}) with $\omega=\tilde\omega=0$. For the fermionic
one, notice that in order to produce our orbifold boundary conditions,
we have to choose $\vec s_0=\vec s_\pi=(0,0,-1)$ and hence must use
$c_0=c_\pi=-p_3$ in Eq.~(\ref{condfer2}).

Let us next study supersymmetry of this action.  The supersymmetry
variation of the bulk action is now given by
\be
\delta \mathcal S_{\rm bk}^0=0\,,\qquad
\delta \mathcal S_{\rm bk}^m=-2\,p_3M\left[\delta(y)-\delta(y-\pi)\right]
\bar\epsilon\bar \Phi  \gamma^5 \sigma_3 \Psi\,,
\ee
while the boundary piece varies into
\be
\delta \mathcal S_{\rm bd}= 2 i  \left[N_0\delta(y)-N_\pi\delta(y-\pi)\right]
\bar\epsilon\bar \Phi\Psi\,.
\ee
Making use of our parity assignments Eq.~(\ref{BCfermionorb}) we
conclude that for these two pieces to cancel we must have
\be
n_0=n_\pi=-p_3.
\ee
To compare with the interval approach, we note again that there
$c_0=c_\pi=-p_3$ and thus we find that for the action to be
supersymmetric, relation (\ref{ecuacion}) must hold. Therefore
departure from the supersymmetric relation (\ref{ecuacion}) implies
supersymmetry breaking.  However in contrast to the interval case
where the action itself is supersymmetric for any values of $N_f$, the
breaking here is explict and can be viewed as coming from localized
soft masses for the even hyperscalars. Splitting the masses $N_f$ into
a supersymmetric and a soft piece, $N_f=-p_3 M+M_f$ we can write the
localized soft breaking Lagrangian as
\be
\mathcal S_{\rm soft}^{\rm hyper}=\int  
 \biggl(M_0 \delta(y)-M_\pi \delta(y-\pi)\biggr)
 \bar\Phi
\Phi\,.
\label{soft}
\ee

Supersymmetry breaking produced by the soft mass terms for even
scalars in the action (\ref{soft}) bears strong similarities with the
usual Scherk-Schwarz supersymmetry breaking by twisted boundary
conditions in the gaugino (and gravitino) sector. In fact twisted
Scherk-Schwarz boundary conditions for the gauginos $\lambda^i$
($i=1,2$) can be produced by localized gaugino soft masses with an
action~\cite{Bagger:2001ep,Delgado:2002xf,vonGersdorff:2004eq}
\be \mathcal S_{\rm soft}^{\rm gauge}=\int \bar\lambda \left(M_0
\delta(y)-M_\pi \delta(y-\pi)\right)\lambda+{\rm h.c.}
\label{softfer}
\ee
However the nature of supersymmetry breaking by boundary scalar masses
is very different that of the Scherk-Schwarz supersymmetry breaking
(which provides a supersoft or finite breaking) as we will see in the
next section.

\section{\sc Supersymmetry breaking by boundary masses}
\label{susybk}

In this section we will study the nature of supersymmetry breaking by
localized scalar masses as in Eq.~(\ref{actionbd}) for 
\be
N_f=\vec p\cdot\vec s_f M+M_f 
\label{split}
\ee
with $M_f\neq 0$. For simplicity we will assume the case of
vector-like fermions analyzed in subsection~\ref{susybreak},
$\omega=\tilde\omega=1/2$, ($R_0=R_\pi=R$) when there are no
supersymmetric masses,~i.e. $M=0$. This case gives rise to a tachyonic
zero mode in the bosonic spectrum for $M_0=M_\pi$ and it is
particularly interesting.

The gauge interactions of the hypermultiplet $\Phi$ depend on the
gauge group. For simplicity we will assume in this section a $U(1)$
gauge group with generator $\mathcal Q$. Consistency with the reality
condition (\ref{reality}) implies that the generator $\mathcal Q$
satisfies~\cite{Zucker:2003qv}
\begin{equation}
\sigma_2 \mathcal Q=-\mathcal Q^{\star}\sigma_2\ ,
\label{realcond}
\end{equation}
and we will then make the choice
\begin{equation}
\mathcal Q=\frac{1}{2}\sigma_3\ .
\end{equation}

The quartic Lagrangian comes from the integration of the $U(1)$
auxiliary field $\vec X$ in
\begin{equation}
\mathcal L_D=2 \vec X^{\,2}+g_5 \bar \Phi\vec\sigma_R\cdot \vec
X\otimes \mathcal Q\Phi
\end{equation}
where $g_5$ is the 5D $U(1)$ gauge coupling and $\vec\sigma_R$ are the
$SU(2)_R$ generators.  Integration of $\vec X$ yields
\begin{equation}
\mathcal L_D=-\frac{1}{8}\, g_5^2(\bar\Phi\vec\sigma_R\otimes\mathcal
Q\Phi)^2\ .
\end{equation}

We will now call the independent components of $\Phi$ as
\begin{equation}
\Phi^1_1=H_1,\quad \Phi^1_2=H_2
\end{equation}
and will use the reality conditions (\ref{reality}) for the other components,
\begin{equation}
\Phi^2_2=\bar H_1,\quad \Phi^2_1=-\bar H_2\ .
\end{equation}
The quartic potential is then given by
\begin{equation}
V_D=\frac{1}{8}\, g^2_5\left(|H_1|^2+|H_2|^2\right)^2\ .
\label{quartic}
\end{equation}

Unlike in section~\ref{eom} we will consider the term in
Eq.~(\ref{actionbd}) as a perturbation and solve the equations of
motion in the absence of it. The mass eigenvalues are then given as
$\Omega_n=n$ and the mass eigenstates can be read off from
Eq.~(\ref{solbos}) by just putting $N_0=N_\pi=0$ there,~i.e.
\begin{align}
H_1=&\frac{1}{\sqrt{2 \pi}}H_1^{(0)}(x)+ \frac{1}{\sqrt{\pi}}\sum_{n=1}^{\infty}\cos ny\
H_1^{(n)}(x)\nonumber\\
H_2=&\frac{1}{\sqrt{\pi}}\sum_{n=1}^{\infty}\sin ny\
H_2^{(n)}(x) \ .
\label{KKd}
\end{align}
Now using the mode decomposition (\ref{KKd}) we can write the boundary
Lagrangian (\ref{actionbd}) as
\begin{equation}
\mathcal L_{bd}=\frac{1}{\pi}\sum_{m,n=-\infty}^{\infty}\left[
M_0-(-1)^{m+n}M_\pi\right]\bar H_1^{(m)}H_1^{(n)} \ .
\label{frontera}
\end{equation}

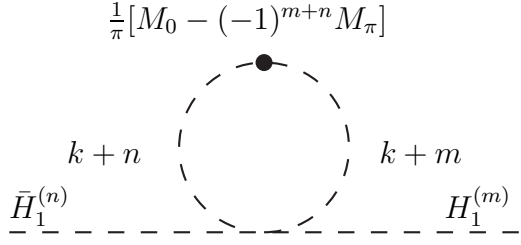
\begin{figure}[htb]
\vspace{1cm}
\begin{center}
\SetScale{1.6}
\begin{picture}(120,60)(0,0)
\DashLine(0,0)(120,0){4}
\DashCArc(60,20)(20,-90,250){4}         
\Vertex(60,40){2}
\Text(90,80)[c]{$\frac{1}{\pi}[M_0-(-1)^{m+n}M_\pi]$}
\Text(0,10)[l]{$\bar H_1^{(n)}$}
\Text(190,10)[r]{$H_1^{(m)}$}
\Text(50,30)[r]{$k+n$}
\Text(140,30)[l]{$k+m$}
\end{picture}
\end{center}
\caption{One-loop diagram renormalizing $M_f$.}
\label{diagrama}
\end{figure}

The renormalization of the boundary mass parameters $M_f$ is given by
loop diagrams induced by the quartic Lagrangian (\ref{quartic}) with
one or more $M_f$-insertions~\footnote{In this toy model the diagram
  with zero mass insertions will be quadratically divergent due to the
  generation of a localized Fayet-Iliopoulos (FI)
  term~\cite{Ghilencea:2001bw}. This can be seen as a renormalization
  of the supersymmetric mass term $M$ and is clearly separable from
  the renormalization of the soft mass terms $M_f$. Of course one
  could avoid the generation of such terms by considering a second
  Higgs which does not interfere with the EW symmetry breaking process
  (as~e.g. a second hypermultiplet with a (large) positive squared
  mass zero mode).}.  Since the leading divergence is given by
diagrams with one mass insertion, we will concentrate in diagrams as
those in Fig.~\ref{diagrama}.

The contribution from the diagrams in Fig.~\ref{diagrama} is
proportional to the factor
\begin{equation}
I=\frac{1}{\pi}\left[M_0-(-1)^{m+n}M_\pi\right]g^2 \mathcal J
\label{factor}
\end{equation}
where $g=g_5/\sqrt{\pi}$ is the 4D gauge coupling, $\mathcal J$ is
given by the Feynman integral
\begin{align}
\mathcal J =&\sum_{k=-\infty}^\infty\int\frac{d^4 p}{(2\pi)^4}
\frac{1}{p^2+(k+m)^2}\, \frac{1}{p^2+(k+n)^2}\nonumber\\
=&\sum_{\ell=-\infty}^\infty\int\frac{d^4 p\, dz}{(2\pi)^4}
\frac{1}{p^2+(z+m)^2}\, \frac{1}{p^2+(z+n)^2}e^{2 i \pi\ell z}
\label{poisson}
\end{align}
and we have made use of Poisson resummation.

The propagators in (\ref{poisson}) have poles in the complex $z$-plane
at locations $z=-n\pm ip$ and $z=-m\pm ip$. In this way for $\ell\neq
0$ the $z$-integrations contour can be closed by an infinite
semicircle. Picking the residues of the corresponding poles provides
the factor
$$e^{-2\pi |\ell|p}$$
that makes the integrand in the remaining
integral to exponentially converge in the limit $p\to\infty$ and the
corresponding integral to be finite. However for $\ell=0$ there
appears a linear divergence. In fact one can write
\begin{align}
\mathcal J=&\int\frac{d^4 p\, dz}{(2\pi)^4}
\frac{1}{p^2+(z+m)^2}\, \frac{1}{p^2+(z+n)^2}+{\rm finite\ terms}\nonumber\\
=&\,\frac{1}{64\pi}\,\Lambda+{\rm finite\ terms}
\label{linear}
\end{align}
where $\Lambda$ is the ultraviolet (UV) cutoff. One can interpret the
result in (\ref{linear}) as a linear renormalization of the brane mass
terms as
\begin{equation}
N_f=M_f\,(1+\Delta),\quad \Delta=\frac{g^2}{64\pi}\,\Lambda R+\cdots
\label{cor}
\end{equation}
Notice that to leading order the radiative corrections to the boundary
mass terms $\Delta$ are boundary independent. Therefore the condition
$M_0=M_\pi$ is not spoiled by the (leading) correction in (\ref{cor}).

Now that we have the loop-corrected localized soft masses one can go
back to section~\ref{eom} and compute the bosonic spectrum to all
orders in the boundary masses as in subsection~\ref{bosons}. In fact
for the model under consideration ($N_0=N_\pi=N$) the bosonic zero
mode is a tachyon with a mass [see Eq.~(\ref{bosspec})]
\be
m_0^2=-N^2(1+\Delta)^2\ .
\ee

A final comment concerning the UV sensitivity of the tachyonic (Higgs)
mass will help to clarify the nature of the supersymmetry breaking
induced by the boundary bosonic masses. This breaking is soft from the
point of view that it does not induces any cubic counterterm in the 5D
theory. However the mass term renormalizes linearly on the boundary,
which induces in turn a linear renormalization in the Higgs mass. This
renormalization can be simply understood from dimensional analysis
since the operator $M_f\bar\Phi\Phi$, in terms of 4D fields, has
dimension three while the gaugino mass operator in (\ref{softfer}),
$M_f\lambda\lambda$, has dimension four: while the former is linearly
sensitive to the cutoff the latter is not. However this sensitivity
does not destabilizes the Higgs mass for values of the cutoff $\Lambda
R\simlt 10^2$: in fact considering for simplicity the weak coupling,
$g^2/64\pi\sim 2\times 10^{-3}$ and $\Delta\simlt 0.2$.  Finally, in
models with a single Higgs the quadratically divergent FI term is the
dominant effect and we would require a lower cutoff ($\Lambda R\simlt
10$) to keep this effect small.

Finally, we have embedded in this section, and for the sake of
analyzing the supersymmetry breaking induced by boundary scalar
masses, a $U(1)$ gauge theory in the interval. We will consider in the
next section how the whole gauge group $SU(2)\otimes U(1)$ can be
similarly embedded.

\section{\sc Electroweak breaking}
\label{ewsb}

We will now consider the case where the hypermultiplet is a doublet
under the $SU(2)_L$ gauge symmetry. To this end we must generalize the
formalism of the previous sections, where only one hypermultiplet was
considered to one where there are two. The reality condition
(\ref{reality}) is now written as
\be
\bar\Phi^i_\alpha=\epsilon^{ij}\rho_{\alpha\beta}\Phi^{\beta}_j
\label{reality2}
\ee
where the tensor $\rho_{\alpha\beta}$ can be written in the
form~\cite{deWit:1984px}
\be 
\rho={\rm diag}(\epsilon\oplus\epsilon)= {\bf 1}\otimes 
\epsilon\quad 
{\rm or}\quad\rho_{\alpha\beta}=\delta_{\alpha_1\beta_1}
\epsilon_{\alpha_2\beta_2}\,
\label{ro}
\ee 
In particular the reality condition for hyperscalars
$\Phi^\alpha_i=\Phi_i^{\alpha_1,\, \alpha_2}$ is given by
\be \Phi^{\alpha_1,\,2}_2=(\Phi^{\alpha_1,\,1}_1)^*\equiv
\bar\Phi^1_{\alpha_1,\,\,1},\quad
\Phi^{\alpha_1,\,1}_2=
-(\Phi^{\alpha_1,\,2}_1)^*\equiv-\bar\Phi^1_{\alpha_1,\,2}\ee
It is now easy to see that the generators of the symmetry group that
preserve the reality constraint must satisfy
\be
\rho\, T^A = -T^{A*}\rho.
\ee
The largest possible symmetry group is thus generated by 
\be
\{\sigma^2\otimes {\bf 1},\ \sigma^1\otimes\sigma^i,\
\sigma^3\otimes\sigma^i,\ {\bf 1}\otimes \sigma^i\} 
\ee
which is the spinor representation of $SO(5)$. As we will see the BC's
will however break this to a subgroup and so does a nonzero mass term
in the bulk.

The reality constraints for the boundary matrices $S$ and $T$ are
given by Eqs.~(\ref{realityS}) and (\ref{realityR}) with the
substitution $\epsilon_{\alpha\beta}\to\rho_{\alpha\beta}$ where the
operator $\rho$ is defined in Eq.~(\ref{ro}). We thus find the
generalizations 
\be
S^T\rho=-\rho S\qquad {\cal M}^T\rho=-\rho{\cal M}  
\ee
while the constraint (\ref{condT}) remains unchanged.  We conclude that $S$
and $\cal M$ are $so(5)$ valued.
We expect the biggest unbroken subgroup if we choose $S_0\propto
S_\pi\propto {\cal M}$.  In fact all such choices are equivalent and
lead to an $SU(2)\otimes U(1)$ subgroup.  The most convenient one is
to take $S_f\propto{\bf 1}\otimes\sigma^3$ which leads to
$SU(2)\otimes U(1)$ generated by
\be \{\sigma^2\otimes {\bf 1},\ \sigma^1\otimes\sigma^3,\
\sigma^3\otimes\sigma^3,\ {\bf 1}\otimes \sigma^3\} \ee

The formal proof of supersymmetry of the action as well as the
solution to the EOM go along similar lines as those followed in
previous sections. In particular the mode decomposition for bosons and
fermions is that given in Eqs.~(\ref{decompbos}) and
(\ref{decompfer}), respectively. In order to have a vectorlike fermion
spectrum as well as unbroken $SU(2)\otimes U(1)$, we will fix
$S_0=-S_\pi={\bf 1}\otimes\sigma^3$, i.e.~$\tilde\omega=1/2$.  The
remaining freedom we have for the fermionic parameters is
$c_0=-c_\pi=\pm 1$. With these parameters the Higgsino mass is given
by (\ref{fermvecspec}) with $c_0=\pm 1$ which for large $M$ gives for the
lightest mode mass $m=2 M\exp(-\pi MR)$ for $c_0=1$ and $m=M$ for $c_0=-1$
respectively. Note that for $MR\simgt1$ the Higgsino becomes too light
for $c_0=+1$ and one should fix $c_0=-1$ instead.  

We will now consider (for illustrative purposes) the model where we
break supersymmetry by choosing $\omega=1/2$, $N_0=N_\pi\equiv N$.
The mass of the Higgs boson doublet is then given by (\ref{bosspec}).
The eigenstate of the (tachyonic) zero mode of the Higgs doublet is
\be
\Phi^{\alpha_1,\,1}_1(x,y)={\cal N}^{-1} e^{-N y} H^{\alpha_1}(x),\qquad
\Phi^{\alpha_1,\,2}_2(x,y)={\cal N}^{-1}e^{-N y} [H^{\alpha_1}(x)]^*
\label{Higgs}
\ee
all other components vanishing.  Here $H(x)$ is the 4D physical Higgs
field and $\Phi$ fulfills the BC's with $S_0=-S_\pi={\bf
1}\otimes\sigma^3$ and $T_0=-T_\pi=-\sigma^3$,
\be
{\displaystyle \frac{1}{2}(1+R_{f})\Phi(x,y_f)=0},\qquad 
R_{f}=-\sigma^3\otimes {\bf 1}\otimes\sigma^3
\label{conphi}
\ee
The normalization factor is determined to be ${\cal N}^2=(1-e^{-2\pi N
R})/2N$.  Notice that $SU(2)_L\otimes U(1)_Y$ acts on the physical
Higgs field $H$ in the standard way, i.e.~by the
generators~\footnote{We normalize the generators to ${\rm tr}
\{T^AT^B\}=\frac{1}{2}\delta^{AB}$.}
$\{\frac{1}{2}\sigma^i,\frac{1}{2}\}.$ The effective 4D theory is
obtained by integrating over the extra dimension.  The mass Lagrangian
becomes
\be
\mathcal L_m=(N^2-M^2)\left|H\right|^2
\label{Higgs2}
\ee

The quartic Lagrangian comes from integrating out the $SU(2)_L\otimes
U(1)_Y$ auxiliary fields $\vec X^A$ where $A=1,\,2,\,3$ labels the
generators of $SU(2)_L$, $T^A$, and $\vec X^4$ the generator of
$U(1)_Y$, $Y$. From the action of the
super-Yang-Mills and hypermultiplets
\be 
\mathcal L_D=2\vec X^A \cdot \vec X^A+g_A
\bar\Phi^j_{\alpha} \left(\vec\sigma_R\right)^i_j \vec X^A
\left(T^A\right)^{\alpha}_{\beta}\Phi^{\beta}_i\ ,
\label{Dlag}
\ee
where $g_A$ is the 5D gauge coupling corresponding to the generator
$T^A$. Integration of $\vec X^A$ in (\ref{Dlag}) yields
\be \mathcal L_D=-\frac{1}{8}\,g^2_A\,\left(\bar\Phi\;
\vec\sigma_R\otimes T^A\;\Phi\right)^2.
\label{Dlag2}
\ee
Next we particularize (\ref{Dlag2}) to the zero mode Higgs
doublet~\footnote{We can assume here that non-zero modes with masses
  controlled by $1/R\simeq$ few TeV are much larger than the weak
  scale and they have been integrated out.} of Eq.~(\ref{Higgs}). We
get the Lagrangian~\footnote{For the $SU(2)_L\otimes U(1)_Y$ group
  with 5D gauge couplings $g_5$ and $g'_5$.}
\be \mathcal
L_D=-\frac{1}{8}\left(g^2_5+g^{\prime\,2}_5\right)
\left|H\right|^4\frac{e^{-4Ny}}{{\cal
N}^2}
\label{Higgs4}
\ee

Putting together Eqs.~(\ref{Higgs2}) and (\ref{Higgs4}), expanding the
neutral component of the Higgs doublet as $H^0=h/\sqrt{2}+i\chi^0$
(where $h$ is the normalized Higgs field with a vacuum expectation
value $\langle h\rangle=v=246$ GeV) and integrating over the fifth
dimension we obtain for the Higgs field the tree-level potential
\be
V=-\frac{1}{2}(N^2-M^2)\,h^2+\frac{1}{32}\left(g^2+g^{\prime\,2}\right)
\,\kappa(\pi NR)\, h^4 \label{Higgs5}\ee
where $g$ and $g'$ are the corresponding 4D gauge
couplings~\footnote{4D and 5D gauge couplings $g_4$ and $g_5$ are
  related to each other as $g_5^2=\pi R g_4^2$.} and $\kappa(\pi NR)$,
defined by
\be
\kappa(x)=x \coth(x)\ ,
\ee
comes from the normalization factor of the zero-mode wave function in
(\ref{Higgs}).  Fixing the minimum of the potential to the physical
value $v$ one finds the tree-level Higgs mass as a function of the
$Z$-boson mass $m_Z$
\begin{align}
&m_H^2=\kappa(\pi NR)\,m_Z^2,\nonumber\\
&N^2-M^2=\frac{1}{2}m_H^2
\label{Higgsm}
\end{align}

\begin{figure}[hbt]
\centering
\epsfig{file=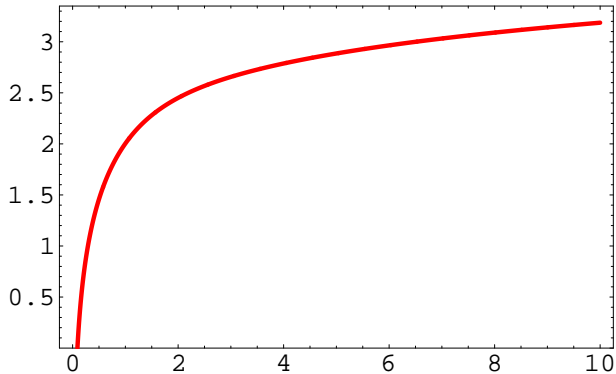}
\caption{\it Plot of $\varepsilon$ as a function of $N$ in TeV. We have
fixed $1/R=4$ TeV}
\label{tuning}
\end{figure}

Some comments about (\ref{Higgsm}) are in order here.  The Higgs mass
in (\ref{Higgsm}) is the tree-level mass. Its natural value is $m_Z$
as $\kappa(x)=1+x^2/3+\dots$ and for values of $N,\, M,\, m_H\simeq
m_Z$, $NR\ll 1$ and $\kappa(\pi NR)\sim 1$, and the second equality in
(\ref{Higgsm}) is naturally satisfied. On the other hand, as in the
minimal supersymmetric standard model (MSSM), to obtain the prediction
of the physical Higgs mass radiative corrections should be added: they
are controlled by top-quark mass and (logarithmically) by the mass and
mixing angle of the third generation squarks. One should meet in this
model the large $\tan\beta$ MSSM prediction for the SM-like Higgs
mass. It seems however possible to enhance the Higgs mass with values
of $N\gg m_Z$. In fact for $\pi NR\simgt 1$ the relation
\be
m_H\simeq m_Z\sqrt{\pi NR}
\ee 
holds. For instance for $NR\simeq 1$, $m_H\simeq 160$ GeV. Of course
the price to pay is that some fine-tuning between $N$ and $M$ is
required from (\ref{Higgsm}). In general a measure of the fine-tuning
$10^{-\varepsilon}$ can be given as
\be 
10^{-\varepsilon}\simeq
\left[m_H^2/N^2\right]=(\pi Rm_Z)^2\
\frac{\coth(\pi NR)}{(\pi NR)} 
\ee
A plot of $\varepsilon$ as a function of $N$ is presented in
Fig.~\ref{tuning}, where we have fixed $1/R\sim 4$ TeV in agreement
with present bounds from electroweak precision
measurements~\cite{Delgado:1999sv}.
In particular an $\sim$1\% tuning implies $N\simeq 1$ TeV and a
tree-level Higgs mass $m_H\simeq 100$ GeV.

Finally one-loop radiative corrections to the Higgs mass will correct
the localized mass $N$ by the factor $1+\Delta$, as it was described
in section~\ref{susybk}. However, as pointed out there, for moderate
values of the cutoff the corrections should be under control and they
will not destabilize the electroweak minimum.

\section{\sc Conclusions}
\label{conclusions}
In this paper we have analyzed the formalism of hypermultiplets
propagating in the five-dimensional interval $[0,\pi R]$. We have
written down an explicit supersymmetric bulk + brane action where the
field boundary conditions are dynamically obtained from the action
principle. The orbifold boundary conditions are obtained as particular
cases. The theory is characterized by three vectors in $SU(2)_R$ space
(two boundary unit vectors $\vec s_f$ and $\vec t_f$ and one bulk unit
vector $\vec p\,$), and one boundary ($n_f$) and one bulk ($M$)
scalars. A misalignment of the vectors $\vec s_f$ on the two
boundaries gives rise to a supersymmetric mass for the hypermultiplet
and that of the vectors $\vec t_f$ is interpreted as the
Scherk-Schwarz supersymmetry breaking. Finally the presence of the
boundary scalars $n_f$ is also a potential source of supersymmetry
breaking if there is a mismatch between $n_f$ and $\vec p\cdot\vec
s_f$. In fact we can define soft scalar masses $M_f$, as $n_f=\vec
p\cdot\vec s_f +M_f/M$, that can break supersymmetry and electroweak
symmetry at the tree level.

While the nature of the Scherk-Schwarz supersymmetry breaking was
already clear, and known to be equivalent to boundary gaugino masses
for 5D vector multiplets, that of supersymmetry breaking by boundary
hyperscalar masses is clearly an issue. In fact while it is known that
the Scherk-Schwarz supersymmetry breaking is one-loop finite, it
provides a two-loop linear divergence corresponding to the one-loop
renormalization of the gauge coupling~\cite{Delgado:2001ex}. We have
proven in this paper that localized hyperscalar masses have one-loop
linear divergences corresponding to the renormalization of a
dimension-three operator on the 4D boundary. As a consequence the
electroweak minimum remains stable for values of the cutoff $\Lambda
R\simlt 10^2$ which means that it does not spoil the little hierarchy.

For the particular example we have worked out in some detail, where
supersymmetry and electroweak breaking are triggered at the tree
level, the natural tree-level value of the Higgs mass is $m_Z$ unless
a fine tuning of parameters is done in which case it can be raised to
somewhat higher values.  It is worth investigating in the future the
nature and softness of other possible supersymmetry and electroweak
breaking patterns as well as the phenomenology of the models presented
in this paper.

\appendix
\section{\sc Appendix: useful identities}

In this appendix we present some useful identities to deal with 
fields obeying reality constraints.  For fermions as in
Eq.~(\ref{symplectic}), we find the following rules for bilinears:
\be \bar\Psi\,\sigma_H\gamma\,\Omega=
-\alpha(\sigma_H)\bar\Omega\,\sigma_H\gamma\,\Psi\,, \ee where
$\gamma=\{{\bf 1},\gamma^M\}$. and $\alpha(\sigma)$ is defined by
$\alpha(\sigma_0)=+1$ and $\alpha(\sigma_i)=-1$. The reality
properties are given by
\be
(\bar\Psi\,\sigma_H\gamma\,\Omega)^*=
-\alpha(\sigma_H)\bar\Psi\,\sigma_H\gamma\,\Omega\,.
\ee
For scalars as in Eq.~(\ref{reality}), the corresponding transposition
rule is
\be
\bar \Phi\, \sigma_R\otimes\sigma_H \,\Sigma=\alpha(\sigma_R)\alpha(\sigma_H)\,
\bar \Sigma \,\sigma_R\otimes\sigma_H\, \Phi\,,
\ee
while the reality properties read
\be
(\bar \Phi\, \sigma_R\otimes\sigma_H \,\Sigma)^*
=\alpha(\sigma_R)\alpha(\sigma_H)\,
\bar \Phi\,\sigma_R\otimes\sigma_H\, \Sigma\,,
\ee

\vspace*{7mm}
\subsection*{\sc Acknowledgments}

\noindent This work was partly supported by CICYT, Spain, under
contracts FPA 2004-02015 and FPA 2002-00748. One of us (GvG) would
like to acknowledge the hospitality of IFAE where part of this work
has been done.

\end{document}